%% file: main.tex
\documentclass[10pt,conference]{IEEEtran}
\IEEEoverridecommandlockouts

\usepackage{cite}
\usepackage{amsmath,amssymb,amsfonts}
\usepackage{hyperref}
\usepackage{algorithmic}
\usepackage{graphicx}
\usepackage{textcomp}
\usepackage{xcolor}
\usepackage{packages}
\def\BibTeX{{\rm B\kern-.05em{\sc i\kern-.025em b}\kern-.08em
    T\kern-.1667em\lower.7ex\hbox{E}\kern-.125emX}}
\begin{document}

\title{Contextual Fairness-Aware Practices in ML: A Cost-Effective Empirical Evaluation}

\author{
    \IEEEauthorblockN{
        Alessandra Parziale\IEEEauthorrefmark{1},
        Gianmario Voria\IEEEauthorrefmark{1},
        Giammaria Giordano\IEEEauthorrefmark{1},
    }
    \IEEEauthorblockN{
        Gemma Catolino\IEEEauthorrefmark{1},
        Gregorio Robles\IEEEauthorrefmark{2},
        Fabio Palomba\IEEEauthorrefmark{1}
    }
    \IEEEauthorblockA{
        \IEEEauthorrefmark{1}Software Engineering (SeSa) Lab, Department of Computer Science - University of Salerno, Salerno, Italy
    }
    \IEEEauthorblockA{
        \IEEEauthorrefmark{2}Universidad Rey Juan Carlos, Madrid, Spain
    }
}

\maketitle

\begin{abstract}
As machine learning (ML) systems become central to critical decision-making, concerns over fairness and potential biases have increased. To address this, the software engineering (SE) field has introduced bias mitigation techniques aimed at enhancing fairness in ML models at various stages. Additionally, recent research suggests that standard ML engineering practices can also improve fairness; these practices, known as fairness-aware practices, have been cataloged across each stage of the ML development life cycle. However, fairness remains context-dependent, with different domains requiring customized solutions. Furthermore, existing specific bias mitigation methods may sometimes degrade model performance, raising ongoing discussions about the trade-offs involved.

In this paper, we empirically investigate fairness-aware practices from two perspectives: contextual and cost-effectiveness. The contextual evaluation explores how these practices perform in various application domains, identifying areas where specific fairness adjustments are particularly effective. The cost-effectiveness evaluation considers the trade-off between fairness improvements and potential performance costs. Our findings provide insights into how context influences the effectiveness of fairness-aware practices. This research aims to guide SE practitioners in selecting practices that achieve fairness with minimal performance costs, supporting the development of ethical ML systems.
\end{abstract}

\begin{IEEEkeywords}
Machine Learning Fairness; Fairness-Aware Practices; Cost-Effectiveness; Empirical Software Engineering.
\end{IEEEkeywords}

\input{sections/intro}
\input{sections/background}
\input{sections/design}
\input{sections/results}

\input{sections/ttv}
\input{sections/conclusion}

\section*{Data Availability}
The data that support the findings of this study are openly available in our online appendix \cite{appendix}.

\section*{Acknowledgment}
We acknowledge the use of ChatGPT-4 to ensure linguistic accuracy and enhance the readability of this article. We acknowledge the support of the European Union - NextGenerationEU through the Italian Ministry of University and Research, Project PRIN 2022 PNRR ``FRINGE: context-aware FaiRness engineerING in complex software systEms" (grant n. P2022553SL, CUP: D53D23017340001) and Project PRIN 2022 ``QualAI: Continuous Quality Improvement of AI-based Systems'' (grant n. 2022B3BP5S, CUP: H53D23003510006).

\balance
\bibliographystyle{IEEEtran}
\bibliography{references}

\end{document}

%% file: sections/intro.tex
\section{Introduction}
Machine Learning (ML) applications continue to spread in diverse contexts, becoming integral in business operations due to their impact on efficiency, decision-making, and innovation \cite{zhou2018human, wang2021comparative, ni2020survey}. However, this trend has raised ethical concerns around \emph{fairness}—the expectation that models make unbiased decisions \cite{mehrabi2021survey}. Bias in training data can lead models to make unfair decisions, presenting ethical and legal risks \cite{pagano2023bias, pessach2022review}.

In response to these challenges, the software engineering (SE) research community, particularly within \emph{SE for Artificial Intelligence}, has proposed multiple bias mitigation techniques, i.e., methods designed to reduce or eliminate biases in machine learning models by operating on data or algorithms. These solutions have been categorized as \emph{pre-processing}, \emph{in-processing}, and \emph{post-processing} \cite{hort2024bias}, based on the ML development stage in which they operate. Different research evaluated these solutions empirically \cite{chen2024fairness, 10.1145/3540250.3549103, hort2021fairea}, demonstrating their efficacy in mitigating bias. As fairness is a \emph{context-dependent} issue, i.e., different ethical concerns arise in different contexts \cite{Fabris_2022}, most of the solutions proposed so far have been evaluated under specific settings, including different application domains. 

Nevertheless, fairness-specific adjustments have implications for the \textit{economic sustainability of businesses}. This dimension refers to the financial sustainability of the software over time \cite{de4966447examining}. In this regard, the application of these bias mitigation solutions may impact operating costs in terms of customer satisfaction, particularly with decreased \textit{performances} of the resulting models \cite{mehrabi2021survey,martinez2022software}. De Martino et al. \cite{de4966447examining} performed a benchmark study of the implications of applying bias mitigation solutions on other sustainability dimensions, such as the economic one, highlighting that applying these algorithms involves complex trade-offs, particularly between fairness and performance. 

Drifting apart from specific bias mitigation methods, recent research in SE highlighted how fairness in ML can be addressed by carefully selecting common engineering practices during the development life cycle \cite{voria2024mapping}. These practices have been defined as \emph{fairness-aware practices}---common practices that have a positive impact on the fairness level of an ML model---and they have been cataloged in the six stages of an engineered ML development life cycle \cite{burkov2020machine}. These start from early stages, with practices like `Multi-objective Optimization' or `Data Balancing' in \emph{`Requirements Elicitation'} and \emph{`Data Preparation'}, to practices like `Model Outcomes Analysis' in the final stage of \emph{`Model Maintenance \& Evolution'}.

Furthermore, these practices have been evaluated through a survey with expert ML developers \cite{voria2024survey}, assessing the extent to which these have a positive impact on fairness, how often they are applied, and the perceived effort to be implemented. Results show that the majority of these practices have a positive impact on fairness, according to practitioners. Still, only a few of them are frequently applied despite not requiring high effort to be implemented \cite{voria2024survey}. Additionally, these practices appear to offer a distinct advantage in managing the fairness-performance trade-off, as they can enhance both fairness and model performance without the potential performance costs sometimes associated with bias mitigation techniques \cite{Raina2022}.

Stemming from these considerations, this work presents an empirical evaluation of fairness-aware practices with two main focuses. First, we perform a \emph{contextual} evaluation, in which we select multiple application domains to understand if specific areas require the application of specific fairness-aware practices. Second, we introduce a novel evaluation metric to perform the trade-off analysis targeting the \textit{cost-effectiveness} \cite{riegg2015cost} of these practices. Such an evaluation is inspired by previous work in SE \cite{PASCARELLA201922} and aims at representing the trade-off between the benefits of implementing a fairness practice in relation to its cost in terms of performance loss.

To conduct our study, we select common datasets presented and used in previous research \cite{Fabris_2022, chen2024fairness, de4966447examining} but spanning over different contexts of application, e.g., Finance with the German credit dataset \cite{german} or Law with the COMPAS dataset \cite{Fabris_2022}. Afterward, we select a set of fairness-aware practices based on insights by practitioners regarding their positive impact on fairness and frequency of application \cite{voria2024survey}. Finally, we perform extensive experimentation with datasets and practices evaluating fairness and performance metrics to establish the cost of mitigating bias fairness through these practices. Results show that fairness-aware practices in ML models vary in effectiveness and cost-effectiveness across domains and datasets. Domains like \textit{Finance} showed significant fairness improvements, while \textit{Economic} showed none. Practices like \textit{Iterative Imputer}, \textit{Simple Imputer}, and \textit{Oversampling} balanced fairness and performance well, while \textit{Mutation Testing} was less effective. These results highlight the importance of context in selecting and evaluating fairness practices.

\textbf{Paper Structure.} Section \ref{sec:background} reviews the research literature relevant to our study, highlighting the key distinctions that enable our work to push the state of the art forward. Section \ref{sec:method} outlines the research questions and describes the method used to address them. In Section \ref{sec:results}, we present and analyze the study’s findings and discuss the implications for both researchers and practitioners. Section \ref{sec:ttv} addresses the primary limitations of the study and the strategies we employed to mitigate them. Lastly, Section \ref{sec:conclusion} offers concluding remarks.

%% file: sections/background.tex
\section{Background and Related Work}
\label{sec:background}
ML fairness, defined as the absence of bias against protected groups in automated decision-making systems \cite{mehrabi2021survey}, has rapidly gained importance in the \textit{Software Engineering} field. This heightened focus is reflected in a diverse and growing body of research that explores fairness from multiple perspectives \cite{pessach2022review, starke2021fairness, chen2024fairness, hort2024bias}. As ethical controversies surrounding ML applications continue to surface \cite{brun2018software,ia_ethical_incidents}, they underscore an urgent need to prioritize fairness in ML practices across the field.

\textbf{Background.} Research has proposed various \textit{bias mitigation approaches} throughout the ML pipeline, categorized as pre-, in-, and post-processing techniques. Pre-processing methods address bias in training data, with solutions like Chakraborty et al.'s \cite{chakraborty2021bias} \textsc{Fair-SMOTE}, a synthetic data augmentation method that preserves model performance, and reweighting techniques by Kamiran and Calders \cite{kamiran2012data}, which adjust instance weights to enhance fairness. In-processing techniques modify learning algorithms to mitigate bias during training; for instance, Zhang et al. \cite{zhang2018mitigating} utilized an adversarial approach, while Chakraborty et al. \cite{chakraborty2020fairway} balanced fairness and performance via multi-objective optimization. Finally, post-processing methods adjust outputs after training to ensure fairness. Galhotra et al. \cite{galhotra2017fairness} proposed \textsc{Themis}, which identifies bias using input perturbation, and Udeshi et al. \cite{udeshi2018automated} introduced \textsc{Aequitas} to enhance bias detection efficiency. Black-box and white-box fairness testing approaches, such as those by Aggarwal et al. \cite{aggarwal2019black} and Zhang et al. \cite{zhang2020white}, employ adversarial sampling to detect and address biases.

\textbf{Related Works.} Recent research has advanced benchmark studies for ML bias mitigation techniques. Hort et al. \cite{hort2021fairea} introduced \textsc{Fairea}, a tool to benchmark bias mitigation methods, focusing on five pre- and in-processing algorithms and non-functional requirements. Chen et al. \cite{chen2023comprehensive} used Fairea in a large-scale study with seven algorithms, finding that mitigation methods can reduce ML accuracy, with effectiveness varying by task, model, protected attribute, and metric set. Zhang and Sun \cite{10.1145/3540250.3549103} adapted ML fairness methods for multiple protected attributes, assessing six algorithms. Recently, Chen et al. \cite{chen2024fairness} benchmarked fairness improvements for multiple protected attributes across eight techniques, while Hort et al. \cite{hort2024bias} proposed a new approach to enhance both fairness and accuracy. Finally, De Martino et al. \cite{de4966447examining} benchmarked bias mitigation algorithms and explored the trade-offs among social sustainability---fairness---, economic sustainability, and environmental sustainability.

\stesummarybox{\faList \hspace{0.05cm} Our Contribution.}{In this work, we empirically evaluate fairness-aware practices and their cost in performance loss. We advance research by (1) evaluating underexplored fairness-aware practices rather than specific bias mitigation techniques, (2) performing a context-dependent and cost-effective evaluation of these solutions, and (3) providing practitioners with practical recommendations on the specific set of fairness-aware practices to apply in their context.}

%% file: sections/design.tex
\begin{figure*}
    \centering
    \includegraphics[width=0.80\linewidth]{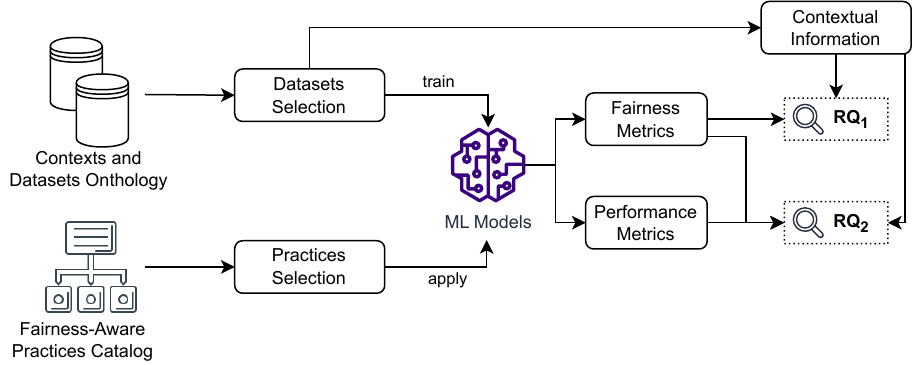}
    \caption{Overview of the research method proposed for our study.}
    \label{fig:method}
\end{figure*}

\section{Research Design and Methods} 
\label{sec:method}
The \emph{goal} of this study is to evaluate the effectiveness of fairness-aware practices in mitigating bias across different contexts, with the \emph{purpose} of assessing their impact and understanding any performance trade-offs. The study considers the \emph{perspective} of both researchers and practitioners. Researchers are interested in the implications of these practices on performance, contributing to the broader discourse on bias mitigation in ML models. Practitioners, meanwhile, seek practical recommendations for embedding fairness-aware practices into their workflows to build fair ML systems.

\subsection{Research Questions} Our empirical study was centered around two main research questions. First, we aimed to quantitatively verify the positive impact of fairness-aware practices on bias mitigation, complementing previous qualitative studies that relied on expert opinions \cite{voria2024survey}. This builds on prior research assessing specific bias mitigation methods \cite{chen2024fairness, de4966447examining}, which reported that practitioners viewed these practices as beneficial for enhancing ML model fairness. We sought to expand on these findings by conducting experiments in diverse contexts to understand the contextual dependency of fairness practices. This led us to our first research question:

\sterqbox{RQ\textsubscript{1}. Contextual Fairness Evaluation.}{To what extent can fairness-aware practices mitigate bias in different contexts?}

Our second objective was to examine the performance trade-offs associated with fairness-aware practices, as the balance between fairness and performance remains a challenge in fairness research \cite{de4966447examining}. To this aim, we designed a cost-effective analysis \cite{riegg2015cost}, i.e., a method for assessing the cost-effectiveness of an intervention by calculating the ratio of its cost to its effectiveness, considering the benefit as fairness gains and the cost as performance loss. This is crucial for both researchers and practitioners, as it can guide cost-effective recommendations for selecting fairness practices that balance fairness and efficiency in ML systems. This motivated our second research question:

\sterqbox{RQ\textsubscript{2}. Cost-Effective Evaluation.}{What is the cost in terms of performance loss against fairness improvements given by the application of fairness-aware practices?}

By exploring these two critical aspects, our study aims to advance fair ML by evaluating practical, easy-to-implement fairness-aware practices in terms of their bias mitigation efficacy across contexts and their performance cost-effectiveness. Figure \ref{fig:method} illustrates an overview of the research approach used to address these questions, with subsequent sections detailing the study objects and methods. Our reporting followed the guidelines of Wohlin et al. \cite{wohlin2012experimentation} and the \textsl{ACM/SIGSOFT Empirical Standards} \cite{empiricalstandards},\footnote{Available at: \url{https://github.com/acmsigsoft/EmpiricalStandards}.} specifically the \textsl{``General Standard''} guidelines given the nature of our study.

\subsection{Practices Selection}
The fairness-aware practices \cite{voria2024mapping} evaluated in this study were selected based on a recent survey of ML experts \cite{voria2024survey}, which assessed these practices' impact on fairness, frequency of application, and perceived implementation effort. By leveraging these findings, we identified a set of practices that offered an optimal balance of high impact, moderate to frequent application, and manageable implementation effort. These practices were deemed suitable for an in-depth, quantitative evaluation, particularly to understand their efficacy in enhancing fairness across various contexts and sensitive attributes. Below, we detail each practice selected, the reasoning behind its inclusion, and the specific implementation choices.

\begin{itemize}

    \item Data Balancing \cite{Valentim2019391}: Experts rated data balancing as a technique with a medium-to-high impact on fairness, achievable with relatively low effort \cite{voria2024survey}. The data balancing techniques selected for implementation include \textit{LabelEncoder}, \textit{Oversampling}, and \textit{Undersampling} due to their popularity \cite{lauron2016improved}.

    \item Parameter Regularization \cite{Vega-Gonzalo2022,Raza2022}: Despite experts identifying parameter regularization as requiring considerable effort to be implemented, it was also noted for its potentially high impact on fairness \cite{voria2024survey}. 

    \item Data Transformation \cite{Biswas2021981}: This practice, frequently utilized according to expert feedback, demands a medium-to-high level of effort but has shown potential for positive impact in different contexts \cite{voria2024survey}. Techniques chosen for implementing data transformation include \textit{IterativeImputer}, \textit{SelectBest}, and \textit{SimpleImputer}.

    \item Metamorphic/Mutation Testing \cite{Ma2020458}: This practice was selected due to its potential positive impact on fairness combined with low implementation effort \cite{voria2024survey}.
\end{itemize}

This selection allowed us to explore practices' impact in diverse settings and provide recommendations for practitioners aiming to build fair ML models with minimal performance trade-offs. As for some practices, different solutions have been proposed, such as Oversampling and Undersampling for Data Balancing, and we ended up with the selection of \textit{eight different techniques}. The code for all the practices is available alongside the experiments in our online appendix \cite{appendix}.

\subsection{Contexts and Datasets Selection}
To conduct a comprehensive evaluation of fairness-aware practices across different application domains, we selected contexts from established fairness research, each represented by a widely used fairness-related dataset \cite{Fabris_2022, le2022survey}. These datasets, chosen for their relevance and prevalence in the field, provide specific contextual settings, enabling us to assess the effectiveness of bias mitigation methods within distinct contexts. Below, we detail the datasets selected, the contexts they represent, and the identified sensitive attributes.

\begin{itemize}
    \item \textsl{COMPAS Dataset}: The COMPAS dataset is a risk assessment tool containing data from 2013 and 2014. It is used to estimate the likelihood of recidivism for defendants and is categorized under the \textit{Recidivism Prediction context} \cite{Fabris_2022}. The sensitive attributes are ``Sex'' and ``Race'', with the disadvantaged groups identified as ``African-American'' for ``Race'' and ``Female'' for ``Sex''.
    \item \textsl{German Credit Dataset} \cite{german}: This dataset is used for credit risk assessment, predicting whether a loan applicant is a good or bad credit risk. It falls under the \textit{Finance context} \cite{Fabris_2022}. The sensitive attributes are ``GenderStatus'' and ``Age'', with disadvantaged groups identified as ``Female-Divorced-Separated-Married'' for ``GenderStatus'' and individuals under the age of 40 for ``Age''.
    \item \textsl{Adult Dataset} \cite{adult}: Created to analyze U.S. population characteristics—such as occupation, education, age, sex, and race—this dataset, derived from census data, aims to predict whether an individual’s income exceeds \$50,000 per year. It represents the \textit{Economics context} \cite{Fabris_2022}. The sensitive attributes in this dataset are ``Race'' and ``Sex'', with the disadvantaged groups being ``Black'' for ``Race'' and ``Female'' for ``Sex''.
    \item \textsl{Bank Marketing Dataset} \cite{bank}: This dataset comprises information on direct marketing campaigns by a Portuguese bank from 2008 to 2013, with the objective of predicting whether a client would subscribe to a bank deposit. It falls within the \textit{Marketing context} \cite{Fabris_2022}. The sensitive attributes are ``Marital Status'' and ``Age'', with disadvantaged groups being ``Married'' for ``Marital Status'' and individuals under 40 for ``Age''.
    \item \textsl{Communities and Crime Dataset} \cite{crime}: Containing socioeconomic data from 46 U.S. states, this dataset is used to predict the total number of violent crimes (including murder, rape, robbery, and assault). It represents the \textit{Crime context}  \cite{Fabris_2022}, with the sensitive attribute being ``Race'' and the disadvantaged group identified as ``Black''.
\end{itemize}

Each of these datasets serves as a context driver for our evaluation, enabling us to observe the impact of fairness-aware practices in settings specific to ``law" or ``economics and business." For instance, assessing practices on a dataset like COMPAS offers insights into how these practices operate within the law context, particularly the one of recidivism prediction. This approach allows us to provide practical insights into the contextual performance and potential trade-offs of fairness-aware methods across distinct application domains.

\subsection{Data Collection and Analysis}
After selecting fairness-aware practices and fairness-related datasets, we experimented with their combination to answer our research questions. In this preliminary empirical study, we focused on \textit{classification} tasks, as they were the most applied in fairness research \cite{Fabris_2022}. Nonetheless, it is worth noting that the selected datasets may be exploited for other tasks as well, e.g., \textit{clustering} or \textit{anomaly detection} \cite{Fabris_2022}. 
Concerning the \textit{classification} task that we implemented in both our \textbf{RQs}, we leveraged the work by Fabris et al. \cite{Fabris_2022} from which we selected the contexts, as they also provided a classification of fairness-related datasets alongside the specific tasks and models that have been used to evaluate them. Hence, we finally selected the \textit{Random Forest} model, as it was among the most commonly used for the chosen datasets. All the data and code used to perform the experiments and evaluate the results are available in our online appendix \cite{appendix}.

\smallskip
\textbf{RQ\textsubscript{1} --- Contextual Fairness Evaluation.}
To address our first research question, we conducted experiments using Random Forest for the \textit{classification} task. We began by establishing a baseline: for each dataset, we trained the model without applying any fairness-aware practices and evaluated its fairness level. To assess model fairness, we used three widely recognized metrics from fairness research \cite{majumder2023fair}: \textit{Average Odds Difference}, which measures the absolute difference between the rates of correct and incorrect classifications across two groups; \textit{False Discovery Rate Difference}, which indicates disparities in false positive rates between distinct groups; and \textit{Disparate Impact}, defined as the ratio of positive outcomes in the protected group to those in the non-protected group. For each dataset, all the metrics were computed on one single protected attribute, selected by considering the ones that were most frequently evaluated in the literature \cite{Fabris_2022}. We selected ``Sex'' for the COMPAS and Adult datasets, ``Age'' for the Bank Marketing dataset, ``Gender'' for the German credit dataset, and ``Race'' for the Communities and Crime dataset. This process resulted in an initial set of five baseline experiments.

Subsequently, we applied each selected fairness-aware practice individually to each dataset and re-trained the same ML model used for the baseline. For each new model, we recalculated the three fairness metrics, leading to 40 additional experiments. Including the baseline, in our data collection phase, we conducted 45 experiments in total. 

To answer \textbf{RQ\textsubscript{1}} and understand if the application of fairness-aware practices may increase the models' fairness level in different contexts, we analyzed such data comparing the baseline's results with the metrics computed after the application of the practices in different contexts. To verify the significance of these results, we finally applied statistical tests. We first assessed the normality of our data to determine the appropriate statistical methods. Using the Shapiro-Wilk test \cite{gonzalez2019shapiro} with a significance level of $\alpha$ = 0.05, we found that not all the studied datasets conformed to a normal distribution, which led us to apply non-parametric methods. Hence, we used the Wilcoxon signed-rank test \cite{woolson2005wilcoxon} to assess differences between the baselines and experiments involving fairness-aware practices, testing the null hypothesis that no significant differences exist. Given the limited sample sizes, further evaluation was not deemed necessary. The application of the Wilcoxon test allowed us to compute p-values and assess statistical significance directly.

\smallskip
\textbf{RQ\textsubscript{2} --- Cost-Effective Evaluation.} To answer our second research question, we followed the same steps as the first one to collect the data. However, in this case, our objective was to evaluate the model's performance. Hence, we collected standard performance metrics of the trained models such as \emph{Precision}, \emph{Recall}, \emph{F1-score}, and \emph{Accuracy} \cite{ml_metrics}. By collecting these data, we ended up with a dataset of experiments composed of 45 rows: for each of the five datasets, we trained an ML model and collected four performance metrics and three fairness metrics, repeating this experiment nine times---one for the baseline and eight for each fairness-aware practice selected. After computing all these experiments and collecting the metrics, we continued with the data analysis phase. 

In the context of \textbf{RQ\textsubscript{2}}, our objective was to evaluate the performance-fairness trade-off of applying fairness-aware practices. Hence, the data analysis process slightly changed from the first research question. We applied a \textit{cost-effectiveness analysis} \cite{riegg2015cost}, a technique used to evaluate the cost-effectiveness ratio of an intervention by dividing its \textit{costs} by its \textit{effectiveness}. In our case, we evaluated each intervention---the application of a fairness-aware practice to an ML model training---based on its \textit{effectiveness} in enhancing fairness relative to its \textit{cost} in terms of model performance loss. This approach enabled us to quantify the trade-offs between improved fairness and reduced predictive accuracy, helping identify the most efficient techniques for maintaining both fairness and performance. 

For each experiment with a fairness-aware practice application, we calculated two measures: (1) the \textit{cost}, computed as the difference in performance between the baseline \textit{(B)} model---without intervention---and the models with the fairness-aware practices applied \textit{(I)}, and (2) the \textit{effectiveness} measured by calculating the difference in fairness metrics between the models with the fairness intervention and the baseline models.

With these two measures, we computed a \textit{cost-effectiveness} ratio by dividing the performance cost by the effectiveness in improving fairness as follows:
\vspace{-1pt}
\[ \text{Cost-effectiveness} = \frac{\text{Performance}_{\text{\textit{B}}} - \text{Performance}_{\text{\textit{I}}}}{\text{Fairness}_{\text{\textit{I}}} - \text{Fairness}_{\text{\textit{B}}}} \]

We used this metric as a comparative metric, enabling us to identify which fairness-aware practice provided the greatest fairness improvements with the least performance compromise. A value \textit{lower than one} indicated a more cost-effective technique, as they yielded higher fairness gains per unit of performance cost. Finally, for each dataset and fairness-aware practice, we computed the cost-effectiveness ratio for all the possible combinations of fairness and performance metrics. We then aggregated these data by averaging the cost, effectiveness, and cost-effectiveness ratios for each combination of dataset and practice. This approach enabled a high-level comparison, identifying which practices consistently balanced fairness and performance.

%% file: sections/results.tex
\section{Analysis and Discussion of the Results}
\label{sec:results}
In the following sections, we present and discuss the results of the empirical study for each dataset, followed by recommendations based on overall cost-effectiveness. The discussion is arranged according to the corresponding \textbf{RQ}.

\begin{table}[h!]
	\centering
    \caption{Wilcoxon Signed-Rank Test p-values for Fairness Metrics Across Contexts. Significant results (p \textless 0.05) are colored and marked with an asterisk (*)}
    \label{tab:rq1_results}
\begin{tabular}{|l|c|c|c|}
\hline
	\rowcolor{purple}
\textbf{\color{white}Context (Dataset)} & \textbf{\color{white}AOD} & \textbf{\color{white}FDRD} & \textbf{\color{white}DI} \\
\hline
Recidivism (COMPAS)               & \cellcolor{purple!15}0.0391*  & 0.7422  & 0.1953 \\\hline
Economic (Adult)                & 0.0547   & 0.4609  & 0.1484 \\\hline
Marketing (Bank Marketing)       & 0.1670   & \cellcolor{purple!15}0.0156* & \cellcolor{purple!15}0.0156* \\\hline
Finance (German Credit)        & \cellcolor{purple!15}0.0078*  & \cellcolor{purple!15}0.0156* & \cellcolor{purple!15}0.0078* \\\hline
Crime (Communities and Crime) & \cellcolor{purple!15}0.0422* & 0.1077  & 1.0000 \\
\hline
\end{tabular}
\end{table}

\subsection{\textbf{RQ\textsubscript{1}} --- Contextual Fairness Evaluation}
In our \textbf{RQ\textsubscript{1}}, we assessed the significance of fairness improvements achieved through fairness-aware practices applied to ML models across various domains. Table \ref{tab:rq1_results} presents the results of our statistical evaluation. More details are available in our online appendix \cite{appendix}.

A key observation is the variation in fairness improvements depending on the context and the specific fairness metrics. For example, in the \textbf{Recidivism} and \textbf{Crime} contexts, significant improvements were observed only in the \textit{AOD} metric. Since these contexts fall under the broader Social Sciences domain, our findings indicate that fairness-aware practices may not consistently yield high equity improvements in these areas.

Conversely, the \textbf{Marketing} domain showed significant improvements in the \textit{FDRD} and \textit{DI} metrics, while \textbf{Finance} demonstrated the most robust results, with significant improvements across all metrics. In contrast, the \textbf{Economic} domain showed the weakest outcomes, with no metric achieving a statistically significant improvement in fairness. These results suggest that the effectiveness of fairness-aware practices is context-sensitive, with some domains more responsive to fairness interventions than others.

The varied significance levels across metrics and domains underscore the complexity of achieving fairness in ML. The effectiveness of fairness-aware practices appears to depend heavily on the context of the application. This variability highlights the need for comprehensive fairness evaluations that account for multiple fairness metrics and domain-specific factors. A universal approach to fairness may be insufficient; researchers and practitioners should analyze the unique characteristics of each domain to design tailored fairness-aware solutions. Contextual factors must be carefully considered to ensure that fairness-aware ML models address the specific challenges of different real-world scenarios.

\stesummarybox{\faList \hspace{0.05cm} RQ\textsubscript{1} --- Summary of the Results.}{The results of our evaluation show that the effectiveness of fairness-aware practices in ML models varies across application domains and fairness metrics. Domains like \textit{Finance} demonstrated significant improvements in all metrics, while \textit{Economic} showed no improvement. Fairness improvements were more limited in domains like \textit{Recidivism} and \textit{Crime}, suggesting that the success of these practices depends on the specific characteristics of each domain. This highlights the importance of context-specific fairness evaluations in ML.}

\begin{table*}
\footnotesize
\centering
\caption{Average Cost-Effectiveness of Fairness-Aware Practices Across Contexts. An asterisk highlights recommended practices.}
\begin{tabular}{|l|c|c|c|c|c|c|}
\hline
\rowcolor{purple}
\textbf{\color{white}Practice} & \textbf{\color{white}Economic (Adult)} & \textbf{\color{white}Marketing (Bank)} & \textbf{\color{white}Crime (Communities\&Crime) } & \textbf{\color{white}Recidivism (COMPAS)} & \textbf{\color{white}Finance (German)}\\ \hline
IterativeImputer & \cellcolor{purple!15}0.0060* & \cellcolor{purple!15}-0.6635* & \cellcolor{purple!15}0.1914* & \cellcolor{purple!15}0.2873* & \cellcolor{purple!15}-0.0158* \\ \hline
LabelEncoder &  \cellcolor{purple!15}0.8648* & \cellcolor{purple!15}-0.1725* & NA & \cellcolor{purple!15}-0.1250* & \cellcolor{purple!15}-0.0158* \\ \hline
Mutation Testing & \cellcolor{purple!15}-0.1423* & 30.0711 & 2.8356 & \cellcolor{purple!15}0.0914* & -1.8315 \\ \hline
Oversample & \cellcolor{purple!15}0.1338* & NA & \cellcolor{purple!15}0.3125* & \cellcolor{purple!15}0.2464* & \cellcolor{purple!15}-0.0837*\\ \hline
Regularization & 1.5822 & -9.4770 & \cellcolor{purple!15}-0.0239* & \cellcolor{purple!15}-0.5797* & \cellcolor{purple!15}-0.7214* \\ \hline
SelectBest & 3.3612 &\cellcolor{purple!15} 0.7065* & \cellcolor{purple!15}-0.2439* & \cellcolor{purple!15}0.4941* & \cellcolor{purple!15}-0.4439*  \\ \hline
SimpleImputer & \cellcolor{purple!15}0.0060* & NA & \cellcolor{purple!15}0.2162* & \cellcolor{purple!15}0.3241* & \cellcolor{purple!15}-0.0158* \\ \hline
Undersample & \cellcolor{purple!15}0.0072* & -13.9254 & \cellcolor{purple!15}0.5744* & \cellcolor{purple!15}0.2477* & \cellcolor{purple!15}-0.2180* \\ \hline
\end{tabular}
\label{tab:cost_effectiveness}
\end{table*}

\subsection{\textbf{RQ\textsubscript{2}} --- Cost-Effective Evaluation}
In the context of \textbf{RQ\textsubscript{2}}, we assessed the cost-effectiveness of applying fairness-aware practices, where fairness gains were treated as benefits and performance loss as costs. This allowed us to recommend practices based on their relative trade-offs. The results of this analysis are presented in Table \ref{tab:cost_effectiveness}. For each fairness-aware practice and dataset context, we report the average \textit{cost-effectiveness ratio} for each combination of performance and fairness metrics computed. All the specific results used for these evaluations, alongside data and code, are available in our online appendix \cite{appendix}. These values should be interpreted as follows: if the cost-effectiveness value exceeds one or falls below minus one for a specific practice on a specific dataset, the performance loss incurred by that practice outweighs its fairness improvement \cite{riegg2015cost}. In Table \ref{tab:cost_effectiveness}, we have highlighted with colors and marked with an asterisk the values indicating positive cost-effectiveness, meaning the specific practice is \textit{recommended} for that dataset. Values marked as ``NA'' result from errors in computing the cost-effectiveness ratios, such as divisions by zero.

For the \textbf{Economic context}, analyzing the Adult dataset, both \textit{Iterative Imputer} and \textit{Simple Imputer} achieved the lowest cost-effectiveness ratio (0.60), suggesting that these data transformation techniques \cite{Biswas2021981} offer the best balance between fairness and performance. \textit{Label Encoder} (0.8648) and \textit{Oversampling} (0.1338) had cost-effectiveness ratios below one, indicating that they improve fairness with minimal performance loss. Similarly, \textit{Undersampling} (0.0072) also demonstrated a favorable trade-off, with high fairness gains relative to a small performance cost. In contrast, \textit{SelectBest} (3.3612) and \textit{Regularization} (1.5822) exceeded the threshold of one, meaning they provide fairness improvements but with more significant performance costs.

In the \textbf{Marketing context}, \textit{Mutation Testing} (30.0711) showed an extremely high cost-effectiveness ratio, indicating a substantial performance loss with minimal fairness improvement, making it highly unfavorable, alongside \textit{Undersampling} (-13.9254) and \textit{Regularization} (-9.4770). Practices like \textit{Label Encoder} (-0.1725), \textit{Iterative Imputer} (-0.6635), and \textit{SelectBest} (0.7065) displayed ratios close to zero, suggesting that their fairness benefits outweigh their performance costs.

For the \textbf{Crime context}, \textit{Oversampling} (0.3125) and \textit{Undersampling} (0.5744) exhibited low cost-effectiveness ratios, making them effective choices as they provide significant fairness improvements with minimal performance trade-offs. \textit{Iterative Imputer} (0.1914) and \textit{Simple Imputer} (0.2162) also showed favorable ratios, indicating they strike a good balance between fairness gains and minimal performance loss. On the other hand, \textit{Regularization} (-0.0239) had a negative cost-effectiveness, reflecting less desirable trade-offs. As a result, \textit{Oversampling}, \textit{Undersampling}, \textit{Iterative Imputer}, and \textit{Simple Imputer} are the most cost-effective choices for this dataset.

In the \textbf{Recidivism prediction context}, using the COMPAS dataset, all practices showed cost-effectiveness ratios below one, indicating they all provide fairness improvements with relatively low performance costs, making them recommended options. Particularly, \textit{Mutation Testing} (0.0914) stands out for offering one of the highest fairness improvements per unit of performance loss, followed by \textit{Label Encoder} (-0.1250).

Finally, for the \textbf{Finance context} represented by the German Credit dataset, all practices demonstrated good cost-effectiveness, except for \textit{Mutation Testing} (-1.8315). The most effective practices were \textit{Iterative Imputer}, \textit{Simple Imputer}, and \textit{Label Encoder} (-0.0158), which effectively increase fairness without significantly harming performance. \textit{Oversampling} (-0.0837) and \textit{SelectBest} (-0.4439) also showed good results.

This dataset-specific analysis highlights that fairness-aware practices are not universally effective across different contexts. Practices applied at the \textit{Data Preparation} stage tend to perform well, offering fairness improvements with minimal performance costs, while \textit{Mutation Testing} is notably less effective. By focusing on practices with a low cost-effectiveness ratio, we can better guide the selection of fairness-aware techniques that optimize the balance between fairness and performance.

\stesummarybox{\faList \hspace{0.05cm} RQ\textsubscript{2} --- Summary of the Results.}{The cost-effectiveness analysis of fairness-aware practices across different datasets revealed that practices applied at the \textit{Data Preparation} stage generally offer good fairness improvements with minimal performance costs. On the one hand, techniques like \textit{Iterative Imputer}, \textit{Simple Imputer}, \textit{Label Encoder}, \textit{Oversampling}, and \textit{Undersampling} were consistently effective, providing a favorable balance between fairness and performance. On the other hand, \textit{Mutation Testing} was generally less effective, showing high-performance costs with minimal fairness gains. Each dataset showed distinct results, highlighting the importance of context in selecting the most cost-effective practices.}

%% file: sections/ttv.tex
\section{Threats To Validity}
\label{sec:ttv}
This section outlines potential threats to the validity of our empirical study and the mitigation strategies applied.

\textbf{Internal Validity.} Internal validity concerns whether our results genuinely reflect the factors under study. A primary threat is the specific implementation choices made for fairness-aware practices. For instance, choosing a regular oversampling technique over SMOTE \cite{Valentim2019391} could impact results. To mitigate this, we closely examined the definitions of these practices \cite{voria2024mapping} and based our implementation decisions on the original design of the cataloged practices. However, alternative implementations could produce different outcomes, influencing both performance and fairness results.

\textbf{External Validity.} External validity pertains to the generalizability of our findings beyond the study's specific setup. We selected contexts for experimentation grounded in recent research \cite{Fabris_2022}. Furthermore, the datasets chosen for each context are frequently used in fairness studies \cite{de4966447examining,chen2024fairness,majumder2023fair}. Nonetheless, our experimentation may not cover all possible contexts, and further studies are needed to validate the broader applicability of our findings. To support replication and further research, all data and scripts are publicly accessible \cite{appendix}.

\textbf{Construct Validity.} Construct validity reflects how well the study’s measurements align with the constructs being evaluated. A potential threat is the choice of datasets to represent different contexts. To address this, we selected widely used datasets \cite{Fabris_2022} that are pertinent to our focus on fairness-performance trade-offs \cite{chen2024fairness,chakraborty2021bias,majumder2023fair,de4966447examining}. Another consideration is our selection of fairness metrics (AOD, FDRD, DI) and performance metrics, which, though not exhaustive, are widely recognized in the literature as robust fairness measures \cite{majumder2023fair, chen2024fairness}. The ML model used could also have influenced results; we, therefore, employed a well-established model common in fairness research \cite{Fabris_2022,chen2024fairness}. For \textbf{RQ\textsubscript{2}}, we based our conclusions on the established cost-effectiveness framework \cite{riegg2015cost}.

\textbf{Conclusion Validity.} Conclusion validity addresses the reliability of our conclusions. A key threat lies in the statistical test applied to \textbf{RQ\textsubscript{1}}, namely, the Wilcoxon signed-rank test \cite{woolson2005wilcoxon}. This test assumes certain data distribution characteristics, and violating these assumptions could impact results. To mitigate this, we assessed the data distribution using the Shapiro-Wilk test \cite{gonzalez2019shapiro} to check for normality, ensuring we selected the most appropriate test for reliable conclusions.

%% file: sections/conclusion.tex
\section{Conclusion}
\label{sec:conclusion}
This paper empirically examines fairness-aware practices—established ML engineering practices known to impact fairness positively. To achieve this, we conducted a contextual evaluation to assess the significance of fairness improvements achieved by these practices. In this evaluation, we carefully selected high-stakes application domains to explore whether specific areas benefit from particular fairness-aware practices. Second, we performed a cost-effectiveness evaluation, treating performance loss as the cost of applying these practices and fairness improvement as the benefit, assessing this trade-off empirically. Our findings indicate that different contexts may require tailored fairness adjustments, as not all practices proved effective across all domains. Furthermore, the cost-effectiveness evaluation revealed that some practices may not be justifiable in specific contexts due to their performance costs, providing practitioners with a preliminary recommendation of which practice they should apply.

The insights gained from our study set the foundation for our future research agenda. First, we plan to broaden our work by including additional tasks, protected attributes, and context-specific metrics to deepen the evaluation of selected practices. Additionally, we aim to develop a recommender system to guide practitioners, based on our extensive experiments, in selecting optimal combinations of fairness-aware practices to achieve fairness in ML systems.